\documentclass{ws-ijmpa}
\usepackage{cite}
\begin{document}



\title{Physics at COSY}

\author{H. Machner}
\address{Inst. f\"{u}r Kernphysik, FZ J\"{u}lich, 52425 J\"{u}lich, Germany}


\maketitle


\begin{abstract}
The COSY accelerator in J\"{u}lich is presented together with its
internal and external detectors. The physics programme performed
recently is discussed with emphasis on strangeness physics.
\end{abstract}



\section{THE ACCELERATOR COMPLEX}

COSY is a synchrotron with electron cooling at injection energy and
stochastic cooling at higher energies. It provides beam of
unpolarised as well as polarised beams of protons and deuteron up to
$3700$ MeV/c momentum. COSY can be used as a storage ring to supply
internal experiments with beam. The beam can also be stochastically
extracted within time bins ranging from 10 s to several minutes to
external experiments. The emittance of the extracted cooled beam is
only $\epsilon=0.4 \pi$ mm mrad. This allows excellent close to
target tracking. Hence a large fraction of the experimental
programme is devoted to meson production close to threshold.

Here we will concentrate on hadron physics thus leaving out
detectors built for different purposes. These are COSY-11, ANKE and
EDDA internally and TOF and BIG KARL externally. The physics at EDDA
will be presented in the contribution by Hinterberger
\cite{Hinterberger} and is therefore neglected here. COSY-11 and
ANKE are magnetic detectors. The former \cite{COSY11} employs an
accelerator dipole magnet while the latter \cite{ANKE} is a chicane
consisting of three dipoles with the middle one as analysing magnet.
TOF \cite{TOF} is a a huge vacuum vessel with several layers if
scintillators. Time of flight is measured between start detectors in
the target area and the scintillators. The target area detectors are
especially suited for the identification of delayed decays and TOF
is thus a geometry detector. BIG KARL \cite{BIG_KARL,BIG_KARL_Neu}
is a focussing magnetic spectrograph of the 3Q2D-type. Particle
tracks are measured in the focal plane area with packs od MWDC's
followed by scintillator hodoscopes allowing for a time of flight
path of 3.5 m. Additional detectors exist. MOMO \cite{MOMO} measured
the emission vertex of charged particles. The Germanium Wall
\cite{GEM} is a stack of four annular germanium diodes being
position sensitive. It acts as a recoil spectrometer.

\section{STRANGENESS PHYSICS}
One strong item in COSY physics is the study of strangeness
production in various processes in $pp$, $pd$ and $pA$ interactions.
Here we have to concentrate on  a few of these reactions.

\begin{figure}[h]
\begin{center}
\includegraphics[width=8 cm]{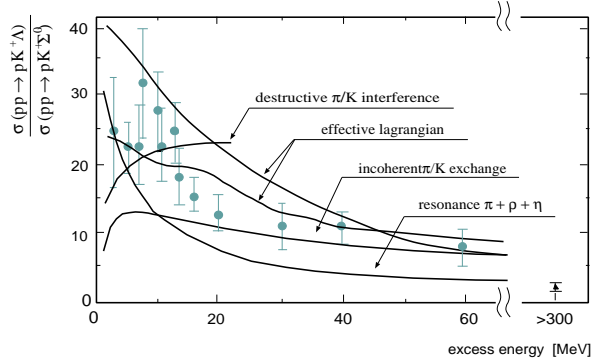}
\caption{Ratio of the cross sections for the indicated associated
strangeness production. The curves are model calculations discussed
in the text.} \label{Associated}
\end{center}
\end{figure}

The $pp\to pK\Lambda(\Sigma)$ reactions, associated strangeness
production were measured by COSY-11 \cite{Bal98,Sew99,Kow02} and TOF
\cite{Marcello01}. In Fig. \ref{Associated} the ratio
$\sigma(pK^+\Lambda)/\sigma(pK^+\Sigma^0)$ is shown as function of
the excess energy.
The ratio rises strongly to threshold. This unexpected behaviour is
studied within several models, including pion and kaon exchange
added coherently with destructive interference \cite{gas99}  or
incoherently~\cite{sib99}, the excitation of nucleon
resonances~\cite{shy01,shy04} (labeled effective lagrangian),
resonaces with heavy meson exchange ($\pi,\rho,\eta$ \cite{sib00}
and heavy meson exchange ($\rho$, $\omega$ and
$K^*$)~\cite{shy01,shy04}. The corresponding curves are also shown
in the figure. All models show a decrease of the ratio with
increasing excitation energy but none of them accounts for all data.

The associated strangeness production is also a useful tool to study
the nucleon--hyperon interaction via FSI. At present a high
resolution study of this interaction runs at BIG KARL. The
measurement of Dalitz diagrams enables even the investigation of the
importance of intermediate $N^*$ excitation.

Connected with the associated strangeness production is the quest
for the existence of a pentaquark. Most of the experimental searches
were performed with electromagnetic probes on the neutron, which, of
course, is embedded in a nucleus. A cleaner environment is the $pp$
interaction.
\begin{figure}[h]
\begin{center}
\includegraphics[width=8 cm]{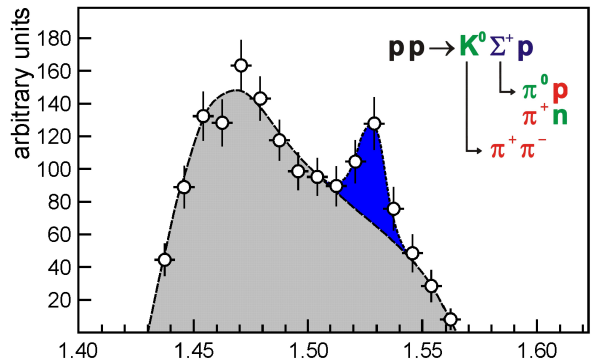}
\caption{Evidence for a pentaquark produced in $pp$ collision.}
\label{Pentaquark}
\end{center}
\end{figure}
The reaction studied with TOF is the
\begin{equation}
pp\to K^0\Sigma^+p
\end{equation}
reaction. The $K^0$ is identified via its decay into two pions and
the $\Sigma$ via its delayed decay. The data are shown in Fig.
\ref{Pentaquark} \cite{Abdel_Bary04}. There is evidence on a
$4\sigma$ level for the production of a pentaquark
\begin{equation}
pp\to \Theta^+\Sigma^+
\end{equation}
with a subsequent decay of the $\Theta^+$ into $K^=$ and $p$. For
the enormous body of papers related to pentaquark we refer to the
talk by Stancu \cite{Stancu04}.

Another intersting reaction is
\begin{equation}
pp\to d K^+\bar K^0.
\end{equation}
On order to reach the threshold of this reaction the maximal energy
of COSY had to be lifted above its design value of 2.5 GeV. The data
were taken at an energy of 2.65 GeV. The analysis of the data
\cite{Grishina04} resulted in a dominance of the channel
\begin{eqnarray}
pp\to d a_0^+
\end{eqnarray}
with a subsequent decay $a_0^+\to K^+\bar K^0$.

At the BIG KARL spectrograph the reaction
\begin{equation}
pd\to ^3HeK^+K^-
\end{equation}
was studied with the MOMO vertex wall.
\begin{figure}
\includegraphics[height=4 cm]{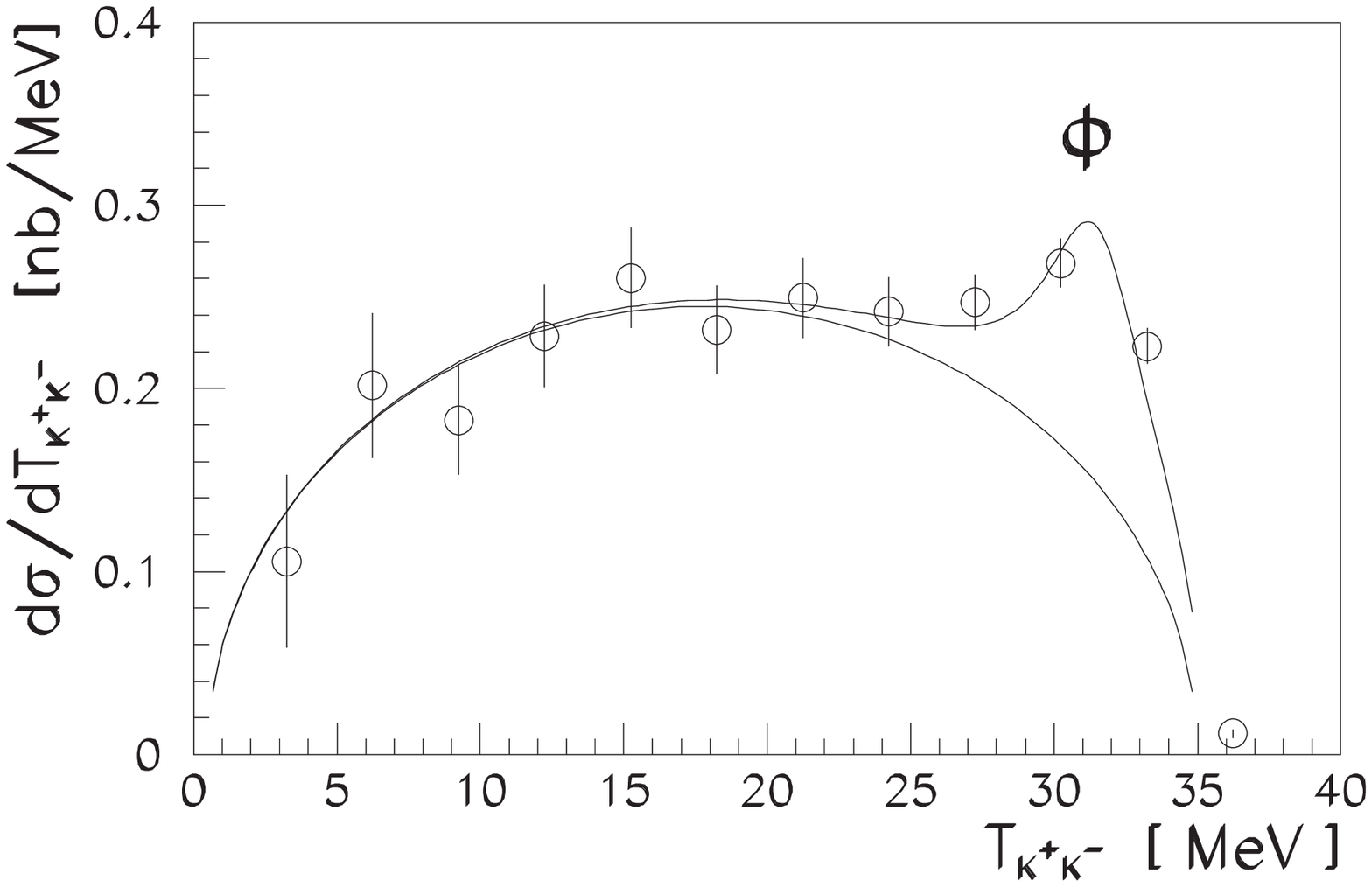}
\includegraphics[height=4 cm]{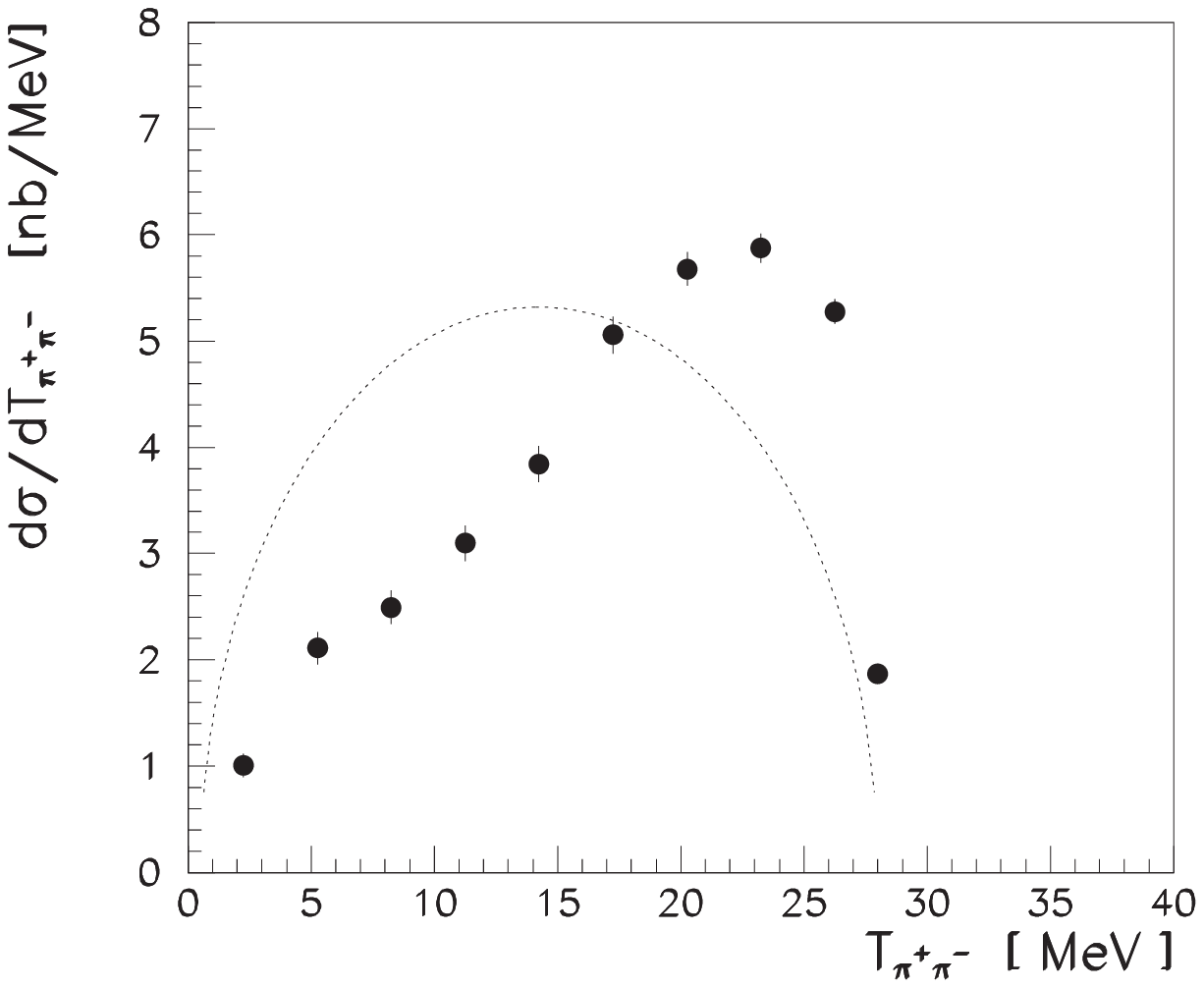}
\caption{Left: Energy spectrum of the two kaons at a maximal excess
energy of 35 MeV. Right: Same as left but for two pions for a
maximal excess energy of 28 MeV.} \label{KK_pipi}
\end{figure}
The interest in this reaction stems from the surprising behaviour of
two pions in the
\begin{equation}
pd\to ^3He\pi^+\pi^-
\end{equation}
reaction \cite{Bellemann99}. The latter reaction showed a p-wave
between the two pions even close to threshold. In Fig. \ref{KK_pipi}
the energy spectrum of two kaons for a maximal energy of 35 MeV are
shown. Besides a smooth continuum the production of $\phi$-mesons is
visible. The energy distribution follows phase space, hence it is
s-wave. The same conclusion  holds for the $KK$--$^3He$ system. Also
the excitation function for the $\phi$ production is an accord with
the assumption of s-wave. To summarise: the $KK$--$^3He$ behaves as
expected while the $\pi\pi$--$^3He$ system shows an unexpected
behaviour. In order to proof the findings for this system the
experiment was repeated but with inverse kinematics. The advantage
of doing so is less settings of the spectrograph. A result of this
measurement is also shown in Fig. \ref{KK_pipi}. It supports the
previous findings.

One aspect of strangeness physics is the $s\bar s$ content in the
nucleon. This is connected to a violation of the OZI-rule in the
ratio
\begin{equation}\label{eqn:OZI}
R=\frac{\sigma(pp\to pp\phi)}{\sigma(pp\to pp\omega)}.
\end{equation}
These two mesons have almost ideal quark mixing and hence the
$\omega$ has negligible $s\bar s$ content while the $\phi$ is an
almost pure $s\bar s$ state (see previous reaction). TOF measured
$\omega$ production at exactly the same excess energy as previous
$\phi$ production, thus allowing the deduction of $R$ as function of
excess energy. This yields $R=(3\pm 1)\times 10^{-2}$ while the
OZI-rule predicts $R=4\times 10^{-3}$. This may point to a serious
content of  $s\bar s$ pairs in the nucleon.

\section{THRESHOLD PRODUCTION, SYMMETRIES, PRECISION EXPERIMENTS}

\begin{figure}
\includegraphics[width=6 cm]{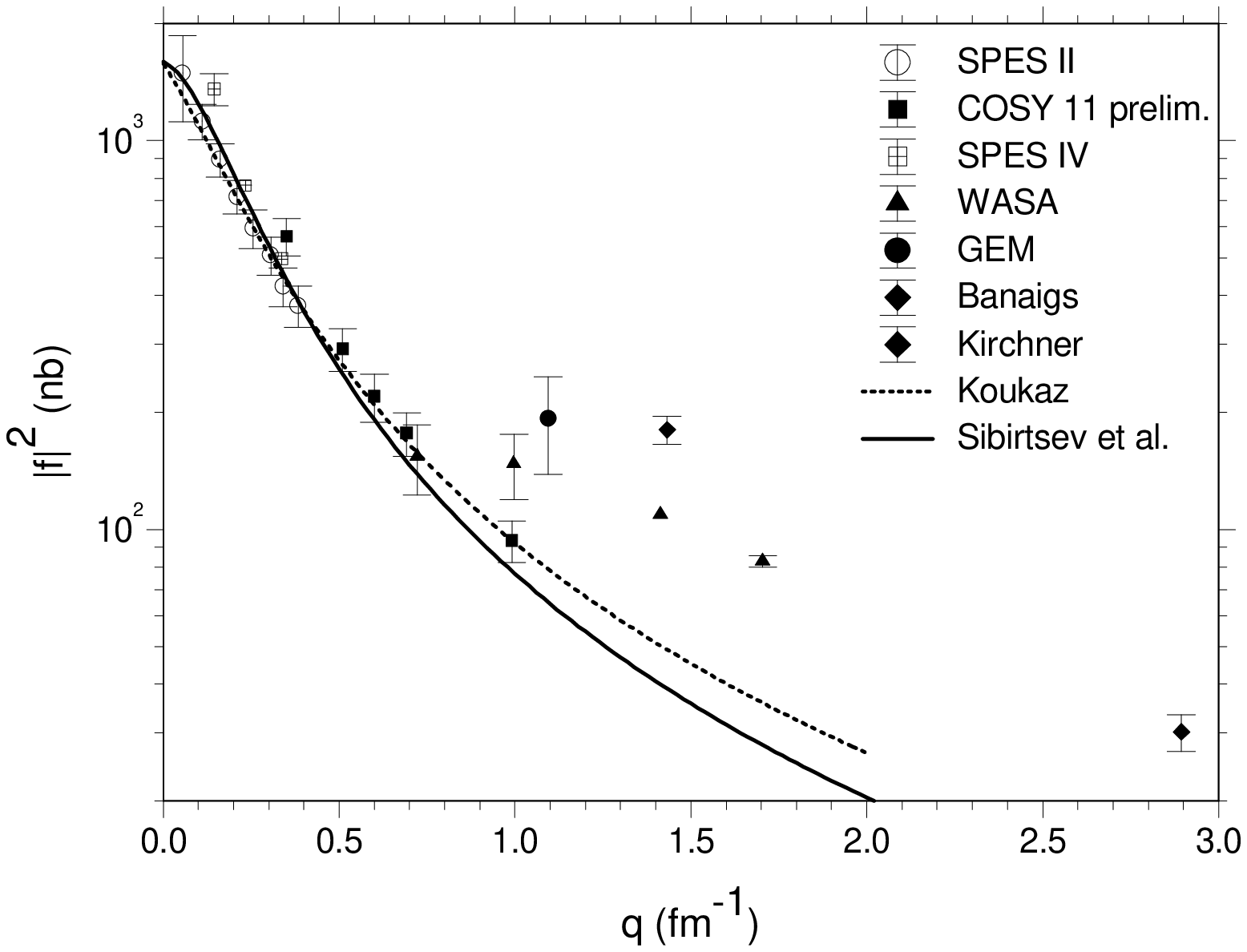}
\includegraphics[width=6 cm]{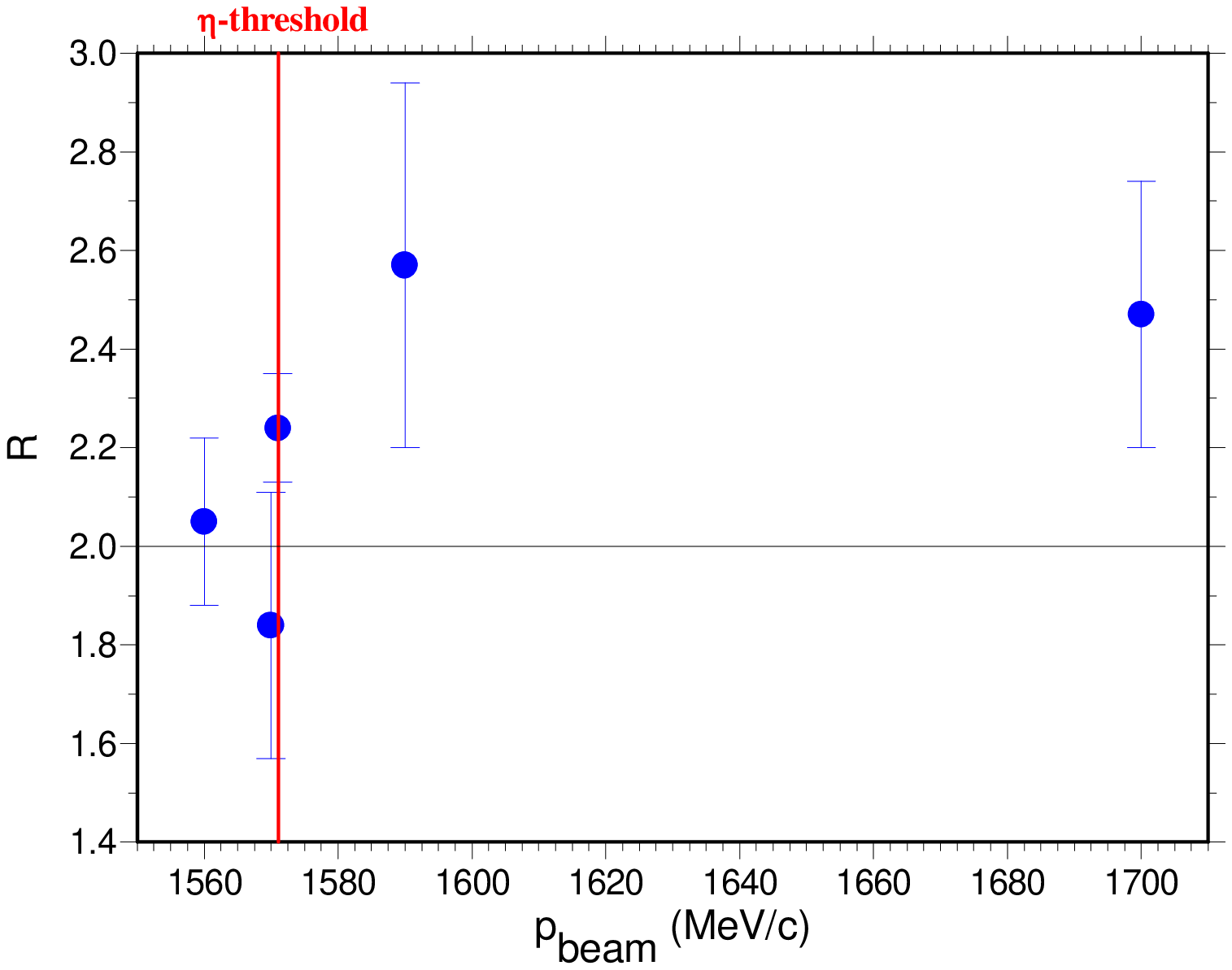}
\caption{Left: Excitation function of the matrix element squared for
the $pd\to {^3He}\eta$ reaction. The references can be found in
\cite{Sibirtsev03}. Right: Excitation functions for the ratio of the
two pion production reactions at maximal momentum transfer (zero
degree in the lab. system). The $\eta$-prodiction threshold is
indicated as line.} \label{iso_threshold}
\end{figure}

There is a wealth of data of light meson production measured at COSY
in the threshold region. The data were taken mainly by the COSY-11
and GEM collaborations for the nucleon-nucleon channel
\cite{Machner99,Moskal_Review,Bettigeri02}. The $\eta$ and $\eta '$
production were discussed by Moskal \cite{Moskal} at this meeting.
Here we will concentrate on $pd\to {^3A}X $ reactions with $A=H$ or
$He$ and $X=\pi,\eta$. The latter reaction is of interest since the
$\eta$-nucleus interaction might be attractive. From the
differential cross sections one can deduce a matrix element $f$ as
\begin{equation}\label{equ:Matrix}
|f\left( \theta  \right)|^2  = \frac{{k^2 }}{{q^2 }}\frac{{d\sigma
}}{{d\Omega }}\left( \theta  \right) = |f_p |^2 |T\left( q
\right)|^2.
\end{equation}
This quantity is shown in the left part of Fig. \ref{iso_threshold}
as function of the of the $\eta$ momentum $q$. Close to threshold
the final state interaction is an s-wave and one can apply the
Watson--Migdal theorem (right side of Eq. \ref{equ:Matrix}) to
extract the FSI amplitude $T\left( q \right)|$ and from this the
$\eta$-${^3He}$ scattering length. Sibirtsev et al.
\cite{Sibirtsev03} performed such an analysis with the result
$a{=}|4.3{\pm}0.3|{+}i(0.5{\pm}0.5)$ fm. This result is also shown in the
figure. The second curve is from A. Khoukaz which includes the
preliminary COSY-11 data.

The GEM collaboration studied isospin symmetry breaking by comparing
neutral and charged pion production in $pp\to d\pi^+$ and $np\to
d\pi^0$ (\cite{BIG_KARL}) reactions, and $pd\to {^3H}\pi^+$ and
$pd\to {^3He}\pi^0$ reactions \cite{Abdel_Samad03}. For the latter
reactions it was found that the angular distribution of the matrix
elements consists of two parts. an exponential part showing scaling
which is attributed to a one step reaction. This part shows isospin
symmetry breaking. The second component is isotropic and is related
to two step processes. It does not show isospin symmetry breaking.
The origin of isospin symmetry breaking is in addition to the
Coulomb force a difference in the masses of the up and down quark.
It was suggested \cite{Magiera00} to study the $pd\to {^3He}\pi^0$
reaction at maximal momentum transfer around the $\eta$-production
threshold. This channel should be sensitive to $\pi^0$-$\eta$ mixing
with the mixing angle being dependent on the different quark masses.
On the contrary the $pd\to {^3H}\pi^+$ reaction should not show such
an interference effect. This was indeed found in an experiment
\cite{Abdel-Bary03} and the ratio of both reactions is shown in Fig.
\ref{iso_threshold}. Baru et al. \cite{Baru03} claimed this effect
to be most probably to FSI. However, if the data are analysed on
terms of the model from Ref. \cite{Magiera00} the mixing angle
results into $\theta=0.006\pm 0.005$. Green and Wycech used a
K-matrix formalism and derived $\theta= 0.010\pm 0.005$. From this
formalism a rather large $\eta$-nucleon scattering length is
extracted making a bound $\eta$-nucleus very likely. The search for
such a system is in progress. Also the question of $\pi^0$-$\eta$
mixing will be further studied via isospin forbidden decays of
$\eta$ and $\eta '$ mesons with WASA at COSY.

\section*{Acknowledgements}

I am grateful to the members of GEM for their collaboration. The
contributions and discussions with Dres. M. B\"{u}scher, A. Gillitzer,
R. Jahn, A. Khoukaz and F. Rathmann is gratefully acknowledged.



\end{document}